\documentclass[doublecol, figures]{epl2} 

\usepackage{latexsym}
\usepackage{amsmath}
\usepackage{amsbsy}

\title{Engineering a spin-fet: spin-orbit phenomena and spin transport induced by a gate electric field}
\shorttitle{Engineering a spin-fet: spin-orbit phenomena induced by a gate electric \dots}

\author{J. L. Cardoso and H. Hern\'andez-Salda\~na}
\shortauthor{J. L. Cardoso and H. Hern\'andez-Salda\~na}

\institute{                    
Area de  F\'{\i}sica Te\'orica y Materia Condensada, UAM-Azcapotzalco,
Avenida San Pablo 180, C\'odigo Postal 02200 M\'exico Distrito Federal,
M\'exico
}
\pacs{85.75.Hh}{Spin polarized field effect transistors}
\pacs{03.65.Ge}{Solutions of wave equations: bound states}
\pacs{71.70.Ej}{Spin-orbit coupling, Zeeman and Stark splitting, Jahn-Teller effect}

\abstract{In this work, we show that a gate electric field, applied in the base of the field-effect devices, leads to inducing spin-orbit interactions (Rashba and linear Dresselhauss) and confines the transport electrons in a two-dimensional electron gas. On the basis of these phenomena we solve analytically the Pauli equation when the Rashba strength and the linear Dresselhaus one are equal, for a tuning value of the gate electric field $\mathcal{E}_g^*$. Using the transfer matrix approach, we provide a joint description of the transport by varying the bias electric field, $\mathcal{E}_b$. We can flip the spin of the incident electrons, or block the spin-down completely. The robustness of this behavior is proved when $\mathcal{E}_g^*$ changes by $\mathcal{E}_g^* \pm \delta \mathcal{E}_g$. 
}

\begin{document}

\maketitle

\section{Introduction} The main consequences of the nanometer-scale devices, with defined boundary conditions, are the confinement of the electric carriers in one-dimensional potential wells and the formation of discrete energy spectra. For example, when an homogeneous magnetic field is applied in a nanostructure, the energy levels are defined by the well-known Landau levels and the associated spectrum can be manipulated by this magnetic field \cite{DCTsui}. For a gate electric field, there are two approaches for studying its influence in a two-dimensional electron gas (2DEG) spectrum: with perturbation methods \cite{GBastard1, PPereyra1, FMAlves} or with Airy functions \cite{DABMiller1, DABMiller2, RdeSousa}. We found the quantization of the transverse energy of the electron, the thickness of the 2DEG and, hence, the induction of the spin-orbit interactions, by taking into account the exactly solvable problem whose solutions are linear combinations of two independent Bessel-$\left( \pm \frac{1}{3} \right)$ functions and simulate the gate junction as an infinity triangular potential well for an idealized Spin-FET, see fig. \ref{well} a) and b).

The manipulation of spin by electric fields in semiconducting environments has generated a lot of theoretical and experimental research aimed at developing useful spintronic devices and novel physical concepts \cite{IZutic, SBandyopadhyay1}. The spin transistor elucidated by Datta and Das\cite{SDatta} is made to drive a modulated spin-polarized current. For this, the spin precession is controlled via the Rashba spin-orbit coupling associated with the interfacial electric fields present in the quantum well that contains the 2DEG. The strength of this interaction can be tuned by the application of an external gate voltage \cite{JNitta, FMireles, JBMiller, YJiang}. The linear Dresselhaus spin-orbit interaction (SOI) could be induced by a gate electric field \cite{RdeSousa, SPrabhakar1, SPrabhakar2, SBandyopadhyay2, SGujarathi}. This work shows how the gate electric field can tune the Rashba and linear Dresselhaus strengths and the interplay between them. For gate electric field value where the interplay is equal to $1$ and its neighborhood, we study the spin transport of the 2DEG in the base by solving analytically the Pauli equation and build the corresponding transfer matrix. 

\section{The 2DEG and the SOI terms induced by a gate electric field} 
\begin{figure}
\centerline{\includegraphics[angle=0,width=2.5in]{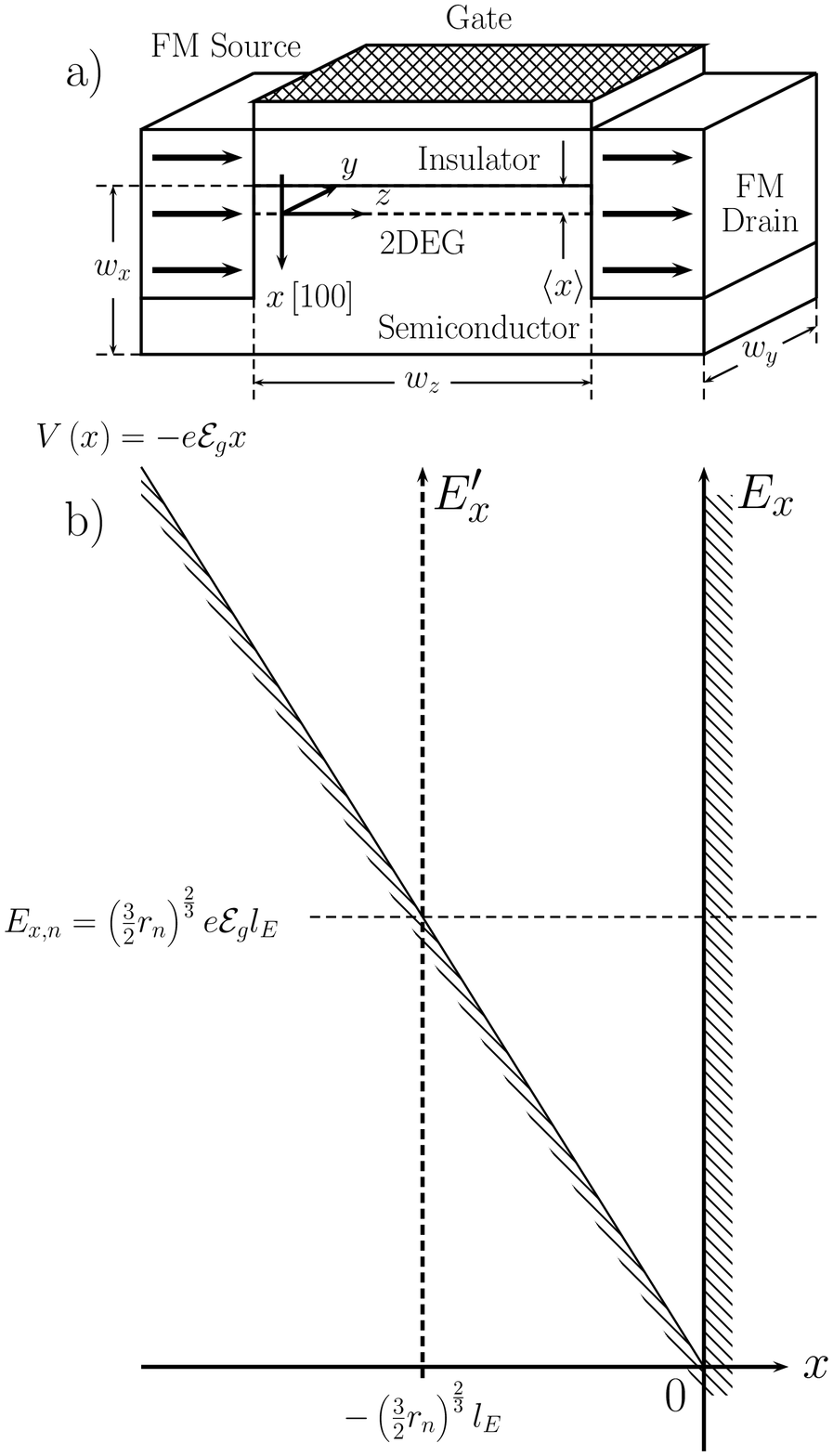}}
\caption{a) An idealized Spin-FET with base's dimensions $w_x$, $w_y$ and $w_z$. We consider here that $\widehat{\imath}$ is oriented along the crystallographic axis $\left[ 100 \right]$. A 2DEG at the plane $yz$ will be formed into the base when a gate electric field is applied and it has a thickness given by the average position $\left\langle x \right\rangle$. b) The gate electric field provokes a confinement due to a triangular potential well $V\left(x\right) = -e\mathcal{E}_gx$ for $-\kappa_z^2 l_{E_g}^3 \leq x \leq 0$. A bound wave-function itself must vanish outside of the triangular well and at the endpoints $x = -\kappa_z^2 l_{E_g}^3 = -\left( \frac{3}{2} r_n \right)^{2/3} l_{E_g}$ and $x = 0$.}
\label{well}
\end{figure}

In this section, we shall study how the gate electric field, $\mathcal{E}_g$, works on the base. It quantizes, with the appropriate boundary conditions, the transverse energy of the electrons, it induces the spin-orbit interactions and the creation the 2DEG. In fig. \ref{well} a), it is shown an idealized spin-FET, with a semiconductor base's dimensions $w_x$, $w_y$ and $w_z$ and two ferromagnetic materials as the source and the drain. We ignore the magnetic field caused by the ferromagnetic contacts. The source-to-drain electric field in the $z$ direction, $\mathcal{E}_b$, is weaker; hence, we neglect its influence to induce any SOI in our approach. 

In order to study the influence of the gate electric field, we take into account the one-dimensional Schr\"odinger equation
\begin{equation} \label{1dSchodinger}
\begin{array}{c}
-\frac{\hbar^2}{2m} \frac{\upd^2 X}{\upd x^2} -e\mathcal{E}_g x X = E_x X
\end{array}
\end{equation}
where $m$ is the effective mass, $e$ is the electron charge and $E_x$ is the transverse energy. It is well-known that the solution of this equation is given by the linear combination of the Bessel-$\left(\pm \frac{1}{3}\right)$ functions $X \left(x \right) = \chi^{\frac{1}{2}} J_{-\frac{1}{3}} \left( \frac{2}{3}\chi^{\frac{3}{2}} \right) A + \chi^{\frac{1}{2}} J_{\frac{1}{3}} \left( \frac{2}{3}\chi^{\frac{3}{2}} \right) B$ with $\chi =  \frac{x}{l_{E_g}} + \kappa_x^2 l_{E_g}^2$, $l_{E_g}^3 = \frac{\hbar^2}{2me \mathcal{E}_g}$ and $\kappa_x^2 = \frac{2m E_x}{\hbar^2}$. Several electrons can be trapped in the base, when the gate electric field is applied. In order to understand the electric field's confinement on the base, we take into account a triangular potential well $V\left(x\right) = -e\mathcal{E}_g x$ formed by $V \left(x\right) = \infty$ for $V\left(x\right) \leq -e\mathcal{E}_gx$ and for $x \geq 0$, as shown in Fig. \ref{well} b), this model is equivalent to the used ones in refs. \cite{DABMiller1, DABMiller2, RdeSousa}. The bound states inside of this triangular potential can be determined if we consider the following boundary conditions $X\left( x = -\kappa_x^2 l_{E_g}^3 \right) = 0$ and $X\left( x =0 \right)= 0$. The most general solution of the one dimensional Schr\"odinger equation, satisfying the first condition, is of the form $X \left(x \right) = \chi^{\frac{1}{2}} J_{\frac{1}{3}} \left( \frac{2}{3}\chi^{\frac{3}{2}} \right) B $. On the other hand, the second boundary condition implies that $\chi = \kappa_x^2 l_{E_g}^2$ and $X \left( 0 \right) = \kappa_x l_{E_g} J_{\frac{1}{3}} \left( \frac{2}{3} \kappa_x^3 l_{E_g}^3 \right) B = 0 $. Hence nontrivial solutions of this problem exist only if $\kappa_x$ has a value such that $\kappa_{x,n} = \left( \frac{3}{2} r_n \right)^\frac{1}{3} \frac{1}{l_{E_g}}$, here $r_n$ is the $n$th-root of the Bessel-$\frac{1}{3}$ function, {\it i. e.} $J_{\frac{1}{3}} \left( r_n \right) = 0$. In this way, the $n$-bound state acquires the energy $E_{x,n} =  \left( \frac{3}{2} r_n \right)^\frac{2}{3} \frac{\hbar^2}{2m l_{E_g}^2} = \left( \frac{3}{2} r_n \right)^\frac{2}{3} e\mathcal{E}_g l_{E_g}$. Geometrically, a bound wave-function itself must vanish outside the triangular well and at the endpoints $x = -\left( \frac{3}{2} r_n \right)^\frac{2}{3} l_{E_g}$ and $x = 0$, see Fig. \ref{well} b). Taking into account a shift $-\left( \frac{3}{2} r_n \right)^\frac{2}{3} l_{E_g}$ of the zero of the $x$-coordinate, we can write the boundary conditions, $X \left(0 \right) = X \left( \left( \frac{3}{2} r_n \right)^{2/3} l_{E_g} \right) = 0$, the one-dimensional Schr\"odinger equation (\ref{1dSchodinger}) is reduced and has the following structure: $ \frac{\upd^2 X}{\upd x^2} +\frac{x}{l_{E_g}^3} X=0$. The average squared momentum can be calculated with the previous equation $\left\langle p_x^2 \right\rangle = \hbar^2 \left\langle \left( -\frac{\upd^2 \ }{\upd x^2} \right) \right\rangle =  \frac{\hbar^2}{l_{E_g}^3} \left\langle x \right\rangle $. By using the boundary conditions $x = 0$ and $x = \left( \frac{3}{2} r_n \right)^{2/3} l_{E_g}$ and considering a Sturm-Liouville problem, we can calculate the average position $\left\langle x \right\rangle$ 
\begin{equation} 
\begin{array}{c}
 \left\langle x \right\rangle = \frac{\int_0^{\left( \frac{3}{2} r_n \right)^\frac{2}{3} l_{E_g}} x^3 \left[ J_{\frac{1}{3}} \left( \frac{2}{3} \left( \frac{x}{l_{E_g}} \right)^{\frac{3}{2}} \right) \right]^2 \upd x}{\int_0^{\left( \frac{3}{2} r_n \right)^\frac{2}{3} l_{E_g} } x^2 \left[ J_{\frac{1}{3}} \left( \frac{2}{3} \left( \frac{x}{l_{E_g}} \right)^{\frac{3}{2}} \right) \right]^2 \upd x } .
\end{array}
\end{equation}
With the change of variable $x = \left( \frac{3}{2} r_n \xi \right)^{2/3} l_{E_g}$, the previous integrals can be written as 
\begin{equation}
\begin{array}{c}
\left\langle x \right\rangle = \left( \frac{3}{2} r_n \right)^{\frac{2}{3}} l_{E_g} \frac{  \int_0^1 \xi^{5/3} \left[ J_{\frac{1}{3}} \left( r_n \xi \right) \right]^2 \upd \xi }{ \int_0^1 \xi \left[ J_{\frac{1}{3}} \left( r_n \xi \right) \right]^2 \upd \xi }.
\end{array}
\end{equation}
The integral of the denominator has a well-known value, $\frac{1}{2} \left[ J_{\frac{4}{3}} \left( r_n \right) \right]^2$, while the integral of the numerator is calculated with the help of the following indefinite integral $\int \xi^{2q+1} \left[ J_{q} \left( \beta \xi \right) \right]^2 \upd \xi = \frac{\xi^{2 (q+1)}}{4q+2} \left\{ \left[ J_{q} \left( \beta \xi \right) \right]^2 + \left[ J_{q+1} \left( \beta \xi \right) \right]^2 \right\}$ when $q = 1/3$ and $\beta = r_n$. In this way, $\left\langle x \right\rangle = \frac{3}{5}  \left( \frac{3}{2} r_n \right)^{\frac{2}{3}} l_{E_g}$ and, therefore, the average squared momentum is written by $\left\langle p_x^2 \right\rangle = \frac{3}{5}  \left( \frac{3}{2} r_n \right)^{\frac{2}{3}} \left( \hbar/l_{E_g} \right)^2$. A 2DEG at the plane $yz$ will be formed into the base when a gate electric field is applied and it has a thickness given by the average position $\left\langle x \right\rangle$. This thickness is numerically equivalent to the one given in \cite{RdeSousa} for the ground state, $r_1$. This thickness decreases monotonically when $\mathcal{E}_g$ grows, but its magnitude increases for higher values of $r_n$; this implies that the 2DEG is created only by small root values $r_n$ and high gate electric fields.

\begin{figure}
\centerline{\includegraphics[angle=0,width=2.5in]{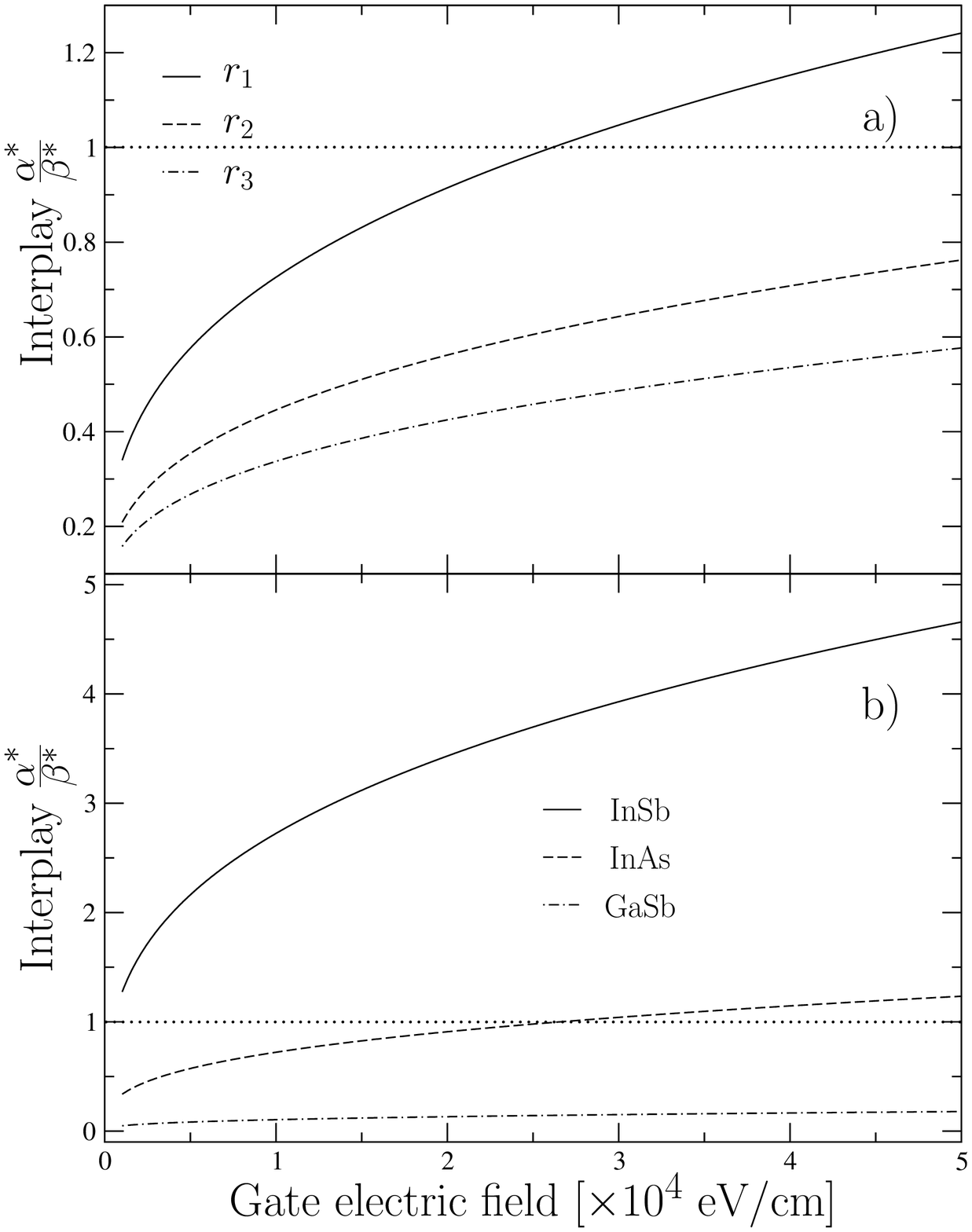}}
\caption{Interplay $\frac{\alpha^*}{\beta^*}$ as a function of the gate electric field. In plot a) GaAs and the roots  $r_1 = 2.9026$, $r_2 = 6.0327$ and $r_3 = 9.1705$. The Dresselhaus strength increases with large values of $r_n$ and the Rashba one does not. When $\frac{\alpha^*}{\beta^*} > 1$ the Rashba SOI dominates the behavior of the 2DEG, while $\frac{\alpha^*}{\beta^*} < 1$ the behavior of the 2DEG is given by the Dresselhaus term. Plot b) shows the interplay $\frac{\alpha^*}{\beta^*}$ as a function of $\mathcal{E}_g$ for three conventional III-V semiconductors InSb, InAs and GaSb at the ground state $r_1$. The dominant SOI in the InSb is the Rashba term, while the GaSb has a strong Dresselhaus one. The SOI terms of the InAs behave similarly as the GaAs does.}
\label{GaAs}
\end{figure}

We consider here that both SOI are induced by the gate electric field $\mathcal{E}_g$. If the geometry of the base is such that $w_x > w_z \geq w_y >> \left\langle x \right\rangle$ for an applied gate electric field, it is possible that the inequalities $\left\langle p_x^2 \right\rangle >> \left\langle p_y^2 \right\rangle_{\mathrm gc} \geq \left\langle p_z^2 \right\rangle_{\mathrm gc} > \left\langle p_x^2 \right\rangle_{\mathrm gc}$ can be true, where the sub-index {\it gc} means geometrical confinement and, for example, $\left\langle p_x^2 \right\rangle_{\mathrm gc} = \left( \pi \hbar/w_x \right)^2$ for the grown state. In the effective-mass approximation, the electron Hamiltonian of a zinc-blendes structure has a spin-dependent $p^3$ coupling called the Dresselhaus term \cite{GDresselhaus}, $\tens{H}_{\mathrm D} = \frac{\beta_{3D}}{\hbar^3} \left[ \tens{\sigma}_x p_x \left( p_y^2 - p_z^2 \right) + \tens{\sigma}_y p_y \left( p_z^2 - p_x^2 \right) \right. \left. +\tens{\sigma}_z p_z \left( p_x^2 - p_y^2 \right) \right]$, where $\beta_{3D}$ is the Dresselhaus three dimensional term, $p_i$ and $\tens{\sigma}_i$ with $i = x$, $y$ and $z$ the components of the momentum and the Pauli matrices respectively. For the triangular potential well and the $x$-direction, which is consider parallel to crystallographic direction $\left[ 100 \right]$, the Dresselhaus term is reduced to $\tens{H}_{\mathrm D} = -\frac{\beta_{3D}}{\hbar^3} \left( \tens{\sigma}_y p_y -  \tens{\sigma}_z p_z\right) p_x^2$. We propose that the most important term of the linear Dresselhaus interaction is written by $\tens{H}_{\mathrm D} = -\frac{\beta^*}{\hbar} \left( \tens{\sigma}_y p_y - \tens{\sigma}_z p_z \right)$, with $\beta^* \simeq \frac{3}{5}  \left( \frac{3}{2} r_n \right)^{\frac{2}{3}} \beta_{3D}/l_{E_g}^2$. For a spin-FET where $w_x \simeq \left\langle x \right\rangle$ the linear Dresselhaus strength could be given by $\beta^* = \beta_{3D} \left[ \frac{3}{5}  \left( \frac{3}{2} r_n \right)^{\frac{2}{3}} 1/l_{E_g}^2 + \left( j \pi/ w_x \right)^2 \right]$; in this work we neglect the geometrical confinement. \cite{JSchliemann, WXu, MOhno}  consider only the geometrical confinement because $w_x << \left\langle x \right\rangle$. On the other hand, the Rashba interaction \cite{EIRashba} has the following form $\tens{H}_{\mathrm R} = \frac{\alpha^*}{\hbar} \left( \tens{\sigma}_y p_z - \tens{\sigma}_z p_y \right)$, with $\alpha^* =a e\mathcal{E}_g$ and $a$ the material's constant. The ratio $\frac{\alpha^*}{\beta^*}$ denotes the interplay between the Rashba strength and the linear Dresselhaus one. Such interplay is given by the following expression
\begin{equation} 
\begin{array}{c}
\frac{\alpha^*}{\beta^*} = \frac{5}{3} \frac{a}{\beta_{3D}} \left( \frac{\hbar^2}{3 r_n m} \right)^\frac{2}{3} \left( e \mathcal{E}_g \right)^\frac{1}{3},
\end{array}
\end{equation}
it is a function of the electric field and depends parametrically on the root $r_n$ and the semiconductor nature ($m$, $a$ and $\beta_{3D}$, table \ref{semicond} has those values). By varying the gate electric field, one can modulate this interplay. Figure \ref{GaAs}a) shows the interplay $\frac{\alpha^*}{\beta^*}$ for GaAs, graphed in the interval $1 \times 10^3 \un{\frac{eV}{cm}} < \mathcal{E}_g < 5 \times 10^4 \un{\frac{eV}{cm}}$ at the roots  $r_1 = 2.9026$, $r_2 = 6.0327$ and $r_3 = 9.1705$. $\frac{\alpha^*}{\beta^*}$ grows monotonically with the gate electric field for a fixed root $r_n$. The horizontal dotted line point up when $\frac{\alpha^*}{\beta^*} = 1$. If we take into account the curve $r_1$ we can find that $\frac{\alpha^*}{\beta^*} = 1$ for a value of the gate electric field, $\mathcal{E}_g^*$, inside the considered interval, but for $r_2$ and $r_3$ these fields are too large. In other words, the Dresselhaus strength increases with large values of $r_n$ and the Rashba does not. When $\frac{\alpha^*}{\beta^*} > 1$ the Rashba SOI dominates the behavior of the 2DEG, while $\frac{\alpha^*}{\beta^*} < 1$ the behavior of the 2DEG is given by the Dresselhaus term. Figure \ref{GaAs}b) shows the interplay $\frac{\alpha^*}{\beta^*}$ as a function of $\mathcal{E}_g$ for three conventional zinc-blende semiconductors InSb, InAs and GaSb at the ground state $r_1$. The curve of the interplay for InSb is over 1, while the corresponding curve of the GaSb is below 1. In this way, the dominant SOI in the InSb is the Rashba term, while the GaSb has a strong Dresselhaus one. The SOI terms of the InAs behave similarly as the GaAs does. Those analitical calculus generalizes the numerical results of ref. \cite{RdeSousa}.

\begin{table}
\caption{Physical properties of the zinc-blende structure semiconductors, according with ref. \cite{RdeSousa}}
\label{semicond}
\begin{center}
\begin{tabular}{l|ccc}
Semiconductor  & $\frac{m}{m_0}$ & $a \left[ \un{\AA^2} \right]$ & $\beta_{3D} \left[ \un{eV \AA{}^3} \right]$ \\
\hline\\
GaAs & 0.0670 & 4.4 & 26 \\
InSb & 0.0136 & 500 & 228 \\
InAs & 0.0239 & 110 & 130 \\
GaSb & 0.0412 & 33 & 187 
\end{tabular}
\end{center}
\end{table}

\section{The transport of the 2DEG under induced spin-orbit interaction}
Into the base, several electrons can be trapped by the gate electric field $\vect{E} =\mathcal{E}_g \widehat{\imath} $ and they form a 2DEG, which is moving in the plane $yz$ under the influence of the induced SOI terms. This movement is described by the Pauli equation with SOI
\begin{equation} \label{fullPE}
\widehat{H} \vect{\Phi} \left( x,y,z \right)  = E_{\textrm F} \vect{\Phi} \left( x,y,z \right)
\end{equation}
where $\widehat{H} = \frac{1}{2m} p^2 -e\mathcal{E}_g x -e\mathcal{E}_b z +V\left(y\right) +\frac{\alpha^*}{\hbar} \left( \tens{\sigma}_y p_z - \tens{\sigma}_z p_y \right) -\frac{\beta^*}{\hbar} \left( \tens{\sigma}_y p_y - \tens{\sigma}_z p_z \right)$ is the Hamiltonian, $\vect{p} = -i\hbar \vect{\nabla}$ is the momentum, $m$ is the effective mass, $\mathcal{E}_b$ is the bias electric field ($\mathcal{E}_b << \mathcal{E}_g$), $E_{\textrm F}$ is the Fermi energy and $\alpha^*$ and $\beta^*$ are the Rashba and the Dresselhaus strengths respectively and $V(y)$ is the transverse confining hard wall potential ($V(y) =0$ for $-\frac{w_y}{2} < y < \frac{w_y}{2}$ and infinite outside this potential region). According to \cite{PPereyra2}, the time reversal operator $\widehat{\mathcal{T}}$ is anti-linear,  {\it i. e.} $\widehat{\mathcal{T}}  \tens{\sigma}_i \to -\tens{\sigma}_i$, and it has the following structure $\widehat{\mathcal{T}} = i \tens{\sigma}_y \widehat{\mathcal{C}}$, with $\widehat{\mathcal{C}}$ the complex-conjugation operator. The Hamiltonian with SOI terms is invariant under time reversal, $\left[ \widehat{\mathcal{T}}, \widehat{H} \right] = 0$. Thus, this Hamiltonian can not produce spontaneous spin polarization \cite{FMireles}. To get rid of the variable $x$, we will consider the complete set of stationary states of a particle in an $x$-dimensional triangular potential well $\left\{ X_n \left(x \right) = \chi_n^{1/2} J_{\frac{1}{3}} \left( \frac{2}{3}\chi_n ^{3/2} \right) B_n \right\}$ with $\chi_n = \frac{x}{l_{E_g}}+\left( \frac{3}{2} r_n \right)^{2/3}$. In other words, we can use these functions to express the wave function $\vect{\Phi} (x,y,z)$ in the form
\begin{equation} \label{trialFunction}
\begin{array}{c}
 \vect{\Phi} \left( x,y,z \right) = \sum_{j=1}^\infty \chi_j^{\frac{1}{2}} J_{\frac{1}{3}} \left( \frac{2}{3}\chi_j ^{\frac{3}{2}} \right) \vect{\phi}_j \left( y,z \right)
\end{array}
\end{equation}
where the expansion coefficients $\vect{\phi}_j \left( y,z \right)$ are spinors. If we introduce this function $ \vect{\Phi} (x,y,z)$ in the Pauli equation (\ref{fullPE}), we have
\begin{equation} \label{2DEG:SOI}
\begin{array}{c}
\left[ \frac{\partial^2}{\partial z^2} +2i \left( \alpha \tens{\sigma}_y +\beta \tens{\sigma}_z\right) \frac{\partial}{\partial z} +\frac{2m E_{\textrm F}}{\hbar^2} - \left( \frac{3}{2} r_n \right)^\frac{2}{3} \frac{1}{l_{E_g}^2} \right. \\
\left. + \frac{z}{l_{E_b}^3} +\frac{\partial^2}{\partial y^2} -2i \left( \beta \tens{\sigma}_y +\alpha \tens{\sigma}_z \right) \frac{\partial}{\partial y}  \right] \vect{\phi}_n \left( x,y \right) =0,
\end{array}
\end{equation}
here $l_{E_b}^3 = \frac{\hbar^2}{2me \mathcal{E}_b}$, $\alpha = m \alpha^*/\hbar^2$ and $\beta = m \beta^*/\hbar^2$. The degrees of freedom of the $x$ axis are completely separated from those in the $yz$ plane. This problem can be considered then as a two-dimensional one and its solutions depend parametrically on $r_n$. We choose the confinement along $x$ to be much stronger, for high gate electric fields, such that only the lowest subband (given by $r_1$) is occupied in this direction under all operating conditions. We only look for solutions around $r_1$, other modes can be solved in a similar way. For clarity, the index $n = 1$ shall be suppressed in the following expressions. In order to get rid of the variable $y$, we will consider the complete set of stationary states of a particle in a $y$-dimensional infinite potential well with boundary conditions $\vect{\iota}_n (-w_y/2) = \vect{\iota}_n (w_y/2) = 0$, which are solutions of the differential equation
\begin{equation} \nonumber
\begin{array}{c}
\frac{d^2 \vect{\iota}}{dy^2} - 2i \left( \beta \tens{\sigma}_y + \alpha \tens{\sigma}_z \right) \frac{d \vect{\iota}}{dy} + q_{y,r} \vect{\iota} =0 .
\end{array}
\end{equation}
It is easy to verify that $\vect{\iota} = \sum_r \vect{\iota}_r = \tens{R}_y \sum_r Y_r \left(y \right) \vect{c}_r$, where $\vect{\iota}$, $\vect{\iota}_r$ and $\vect{c}_r$ are spinors, $Y_r \left(y \right) = \cos \left( k_{y,r} \ y\right) $ for $r$ odd and $Y_r \left(y \right) = \sin \left( k_{y,r} \ y\right)$ for $r$ even, $k_{y,r} = \frac{r \pi}{w_y}$, $q_{y,r} = k_{y,r}^2 +\alpha^2 +\beta^2$ and $\tens{R}_y = e^{i \left( \beta \tens{\sigma}_y + \alpha \tens{\sigma}_z \right)y}$ is a rotation operator. If we use the complete set of functions $\left\{ \vect{\iota}_r \right\}$ to expand $\vect{\phi} \left(y,z\right)$, we get $\vect{\phi} \left(y,z\right) = \tens{R}_y \sum_{r=1}^{\infty} Y_r (y) \vect{Z}_r \left( z \right)$. By introducing this function in the Pauli equation (\ref{2DEG:SOI}), multiplying by $\tens{R}_y^\dagger Y_s \left(y \right)$ and integrating on the variable $y$, we have
\begin{equation} \label{1DEG:SOI}
\begin{array}{c}
\left\{ \frac{\upd^2 }{\upd z^2} +\frac{z}{l_{E_b}^3} +\kappa_{z,s}^2 +\eta^2 +2i \left[ \frac{\alpha \tens{\sigma}_y}{\eta^2} \left( 2\beta^2 -\Delta J_{s,s} \right) \right. \right.\\
\left. \left. +\frac{\beta \tens{\sigma}_z}{\eta^2} \left( 2\alpha^2 + \Delta J_{s,s} \right) \right] \frac{\upd}{\upd z} \right\} \vect{Z}_s \\
+2i\frac{\Delta}{\eta^2} \sum_{r} \left[ \eta \tens{\sigma}_x I_{r,s} -\left( \alpha \tens{\sigma}_y -\beta \tens{\sigma}_z \right) J_{r,s} \right] \frac{\upd \vect{Z}_r}{\upd z} =0
\end{array}
\end{equation}
here $\kappa_{z,s}^2 = \frac{2m E_{\textrm F}}{\hbar^2} -k_{y,s}^2 - \frac{2.6664}{l_{E_g}^2}$, $\eta = \sqrt{\alpha^2 +\beta^2}$, $\Delta = \alpha^2 -\beta^2$, and the mixing terms are given by $I_{r,s} = \frac{2}{w_y} \int_{-w_y/2}^{w_y/2} \sin 2\eta y \ Y_r (y) Y_s (y) \upd y$ and $J_{r,s} = \frac{2}{w_y} \int_{-w_y/2}^{w_y/2} \cos 2\eta y \ Y_s (y) Y_r (y) \upd y$. Notice that if $\beta = \pm \alpha$ the previous equations are uncoupled; in ref. \cite{JSchliemann} a similar case is studied. However, this symmetry would be broken with a magnetic field pointing in any direction. 

According to fig. \ref{GaAs} a) for the GaAs [fig. \ref{GaAs} b) for the InAs], the value of the gate electric field $\mathcal{E}_g^* = 2.6126 \times 10^{4} \un{\frac{eV}{cm}}$ ($\mathcal{E}_g^* = 2.6735 \times 10^{4} \un{\frac{eV}{cm}}$) gives the interplay $\frac{\alpha}{\beta} = 1$. If we take values in the neighborhood of $\mathcal{E}_g^* \pm \delta \mathcal{E}_g$, where $\frac{\delta \mathcal{E}_g}{\mathcal{E}_g^*} = \pm 0.06$, it is possible to show that $\beta = \alpha \left( 1 \pm 0.02 \right)$ and the eq. (\ref{1DEG:SOI}) can be reduced in the first approximation
\begin{equation}
\begin{array}{c}
\left\{ \frac{\upd^2 }{\upd z^2} +\frac{z}{l_{E_b}^3} +\kappa_{z,s}^2 +2\alpha^2 \left(1 \pm 0.02 \right)  \right.\\
\left. +2i \alpha \left[ \tens{\sigma}_y + \left( 1 \pm 0.02 \right) \tens{\sigma}_z \right] \frac{\upd}{\upd z} \right\} \vect{Z}_s \\
\pm i  0.04 \alpha \sum_{r} \left[ \sqrt{2} \tens{\sigma}_x I_{r,s} -\left( \tens{\sigma}_y -\tens{\sigma}_z \right) J_{r,s} \right] \frac{\upd \vect{Z}_r}{\upd z} =0 .
\end{array}
\end{equation}
The coupling terms are not significant and, therefore, they can be negligible and the previous coupled eqs. become the following system of uncoupled eqs. 
\begin{equation}
\begin{array}{c}
\left\{ \frac{\upd^2}{\upd z^2} +2i\alpha \left( \tens{\sigma}_y +\tens{\sigma}_z \right) \frac{\upd}{\upd z}  + \frac{z}{l_{E_b}^3} +\kappa_{z,s}^2 +2\alpha^2 \right\} \vect{Z}_s =0.
\end{array}
\end{equation} 
For each wavenumber $\kappa_{z,s}$, there are two propagating physical channels: one with spin-up and another with spin-down. The general solutions of these equations are given by $\vect{Z}_{s} \left(z \right) = \tens{A}_s \left(z \right) \vect{C}_1 + \tens{B}_s \left(z \right) \vect{C}_2$ being $\tens{A}_s \left( z \right) = e^{-i \alpha l_{E_b} \left( \tens{\sigma}_y + \tens{\sigma}_z \right) \zeta_s} \zeta_s^{\frac{1}{2}} J_{-\frac{1}{3}} \left( \frac{2}{3}\zeta_s^{\frac{3}{2}} \right)$, $\tens{B}_s \left( z \right) = e^{-i \alpha l_{E_b} \left( \tens{\sigma}_y + \tens{\sigma}_z \right) \zeta_s} \zeta_s^{\frac{1}{2}} J_{\frac{1}{3}} \left( \frac{2}{3}\zeta_s^{\frac{3}{2}} \right)$, $\zeta_s = \frac{z}{l_{E_b}} + \left( \kappa_{z,s}^2 + 4\alpha^2 \right) l_{E_b}^2$ and $\vect{C}_1$ and $\vect{C}_2$ are spinors. We now match the wave function $\vect{Z}_{s}$ and its derivative $\vect{Z}_{s}^\prime$ at the borders of the base region. In this way, we can connect the incident waves (to the left-hand side of the source) with the outgoing ones (to the right-hand of the drain). The evolution of the spin-$1/2$ state vectors is governed by the two-channel $4\times4$ transfer matrix
\begin{equation} \label{transm-H}
{\bf M}_{B} \left( z \right) = 
\tens{\Lambda}^{-1} 
\left[ \begin{array}{cc}
\tens{A}_s \left( z \right) & \tens{B}_s \left( z \right)\\
\tens{A}_s^\prime \left( z \right) & \tens{B}_s^\prime \left( z \right) 
\end{array} \right] 
\tens{\Gamma} \tens{\Lambda}
\end{equation}
where 
\begin{equation} \nonumber
\begin{array}{c}
\tens{\Lambda} = 
\left[ \begin{array}{cc} 
\tens{1} & \tens{1} \\
i \kappa_{z,s} \tens{1} & -i \kappa_{z,s} \tens{1}
\end{array} \right], \\
\tens{\Gamma} =
\left[ \begin{array}{cc}
\tens{1} & \tens{0} \\
i \alpha l_{E_b}\left( \tens{\sigma}_y + \tens{\sigma}_z \right) & \tens{1} 
\end{array} \right] ,
\end{array}
\end{equation}
$\tens{A}_s^{\prime}$ and $\tens{B}_s^{\prime}$ are just the derivatives of $\tens{A}_s$ and $\tens{B}_s$ with respect to $z$ and $\tens{1}$ is the unit matrix. This transfer matrix has the symplectic structure\cite{PPereyra2}
\begin{equation} \label{transm-sc}
\tens{M}_{B} =  
\left[ \begin{array}{cc}
\tens{a} & \tens{b} \\
\tens{kb}^*\tens{k}^T & \tens{ka}^*\tens{k}^T
\end{array} \right],
\end{equation}
with $\tens{k} = i\tens{\sigma}_y$. Given $\tens{M}_B$ we are ready to calculate the whole Spin-FET transmission amplitude 
\begin{equation} \label{trans-ncell}
\tens{t} = \left( \tens{a}^{\dagger} \right)^{-1} = \left(
\begin{array}{cc}
t_{\uparrow, \uparrow} & t_{\uparrow, \downarrow} \\
t_{\downarrow, \uparrow} & t_{\downarrow, \downarrow}
\end{array} \right),
\end{equation}
where $t_{i,j}$ (here $i$ and $j$ label the spin-up ($\uparrow$) or spin-down ($\downarrow$) projections) is the transmission amplitude from channel $j$ on the source to channel $i$ on the drain, and $\tens{a}$ is the (1,1) block of the spin-FET transfer matrix $\tens{M}_B$. The off-diagonal terms $t_{\uparrow, \downarrow}$ and $t_{\downarrow, \uparrow}$ are the transmission amplitudes for processes where an odd number of spin flips have taken place inside the base. The corresponding transmission coefficients are defined by $T_{i,j} = \left| t_{i,j} \right|^2$.

\begin{figure}
\centerline{\includegraphics[angle=0,width=2.5in]{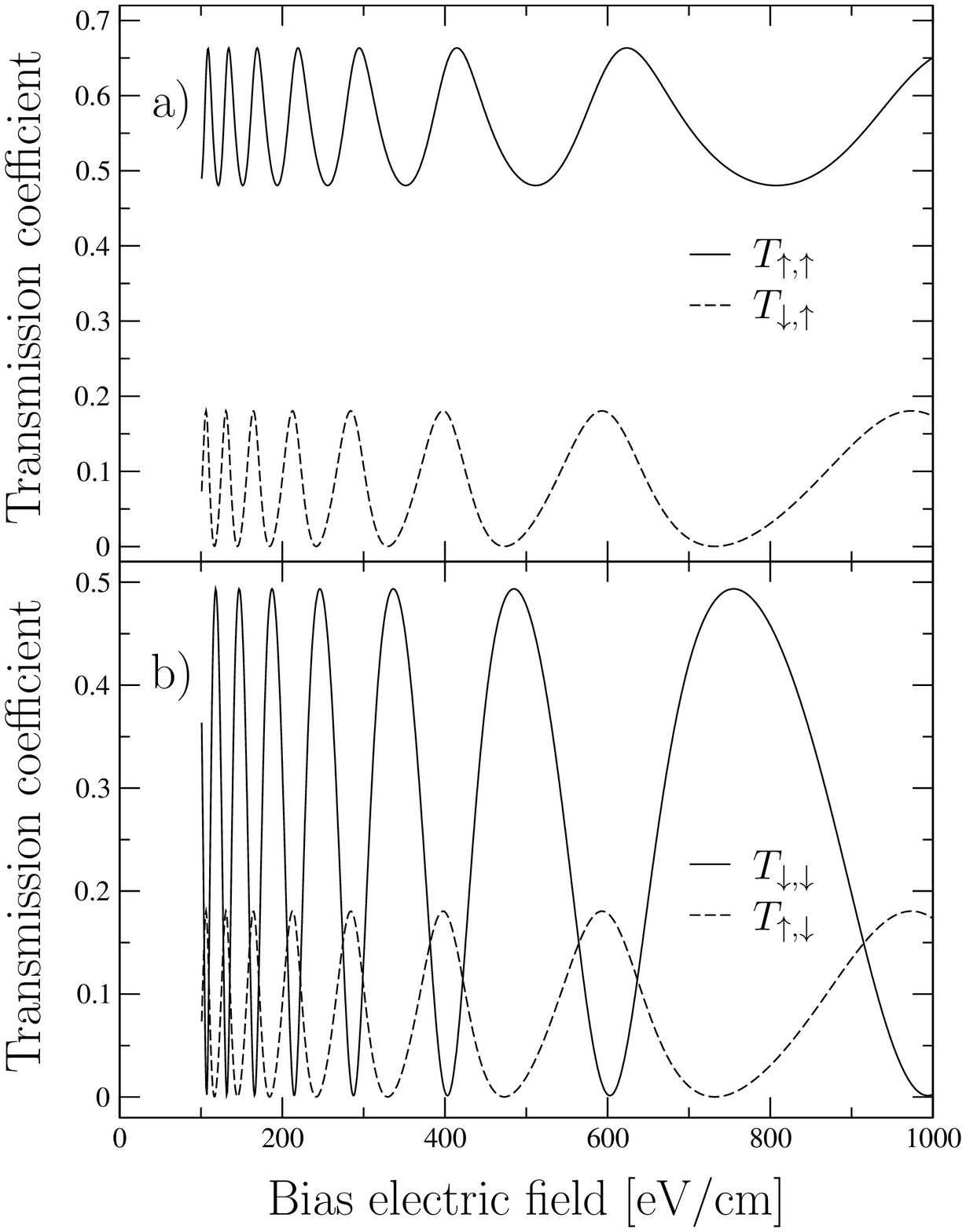}}
\caption{Transmission coefficients as functions of the bias electric field. In the panel a) we have the transmission coefficients $T_{\uparrow, \uparrow}$ and $T_{\downarrow, \uparrow}$, while in the panel b) we have the transmission coefficients $T_{\downarrow, \downarrow}$ and $T_{\uparrow, \downarrow}$.  By varying $\mathcal{E}_b$, the transmission can be modulated more efficiently: we can flip the spin of the incident electrons, or block the spin-down completely, and thus establish a spin transistor.}
\label{transmission}
\end{figure}

We consider electrons of energy $E_{\textrm F} = 78 \un{meV}$ injected into a $\chem{InAs}$  base structure, with $\mathcal{E}_g^* = 2.6735 \un{\frac{eV}{cm}}$ for $\frac{\alpha}{\beta} = 1$. In this report, we will keep fixed the following geometrical variables: $w_x = 5 \times 10^{-1} \un{\mu m}$ and $w_y = w_z = 5 \times 10^{-2} \un{\mu m}$. We are interested here on the visualization of spin-transitions based on the transmission coefficients. We shall start considering the effect of varying the bias electric field on the transmission coefficients, we can then use it as a tuning spin-transition parameter. In the panel a) of fig. \ref{transmission} we have the transmission coefficients $T_{\uparrow, \uparrow}$ and $T_{\downarrow, \uparrow}$, while in the panel b) we have the transmission coefficients $T_{\downarrow, \downarrow}$ and $T_{\uparrow, \downarrow}$.

It is easy to see the influence of the rotation operator $e^{-i \alpha l_{E_b} \left( \tens{\sigma}_y + \tens{\sigma}_z \right) \zeta_s}$, not only on the well known shifting between $T_{\uparrow, \uparrow}$ and $T_{\downarrow, \downarrow}$, but also on the spin transition $\uparrow \longleftrightarrow \downarrow$ processes. The shifting phenomena is given by $-i \alpha l_{E_b} \zeta_s \tens{\sigma}_z$ influence, while the spin transition depends on $-i \alpha l_{E_b} \zeta_s \tens{\sigma}_y$. The general trend of $T_{\uparrow, \uparrow}$ and $T_{\downarrow, \downarrow}$ is oscillatory. The minimum values of $T_{\downarrow, \downarrow}$ are around zero and the minimum values of $T_{\uparrow, \uparrow}$  are not. The transmission coefficients $T_{\uparrow \downarrow}$ and $T_{\downarrow \uparrow}$, exhibit the oscillatory behavior of the two uncoupled-spin transmission coefficients. These maximum and minimum reflect the passage of flux from one spin state to another. The origin of these transitions is the precession term of $e^{-i \alpha l_{E_b} \left( \tens{\sigma}_y + \tens{\sigma}_z \right) \zeta_s}$, which stimulates a mixing between the propagating physical channels. The minimum values of $T_{\uparrow \downarrow}$ and $T_{\downarrow \uparrow}$ are close to zero too. Here, we show that by varying $\mathcal{E}_b$, the transmission can be modulated more efficiently: we can flip the spin of the incident electrons, or block the spin-down completely, and thus establish a spin transistor.

\begin{figure}
\centerline{\includegraphics[angle=0,width=2.5in]{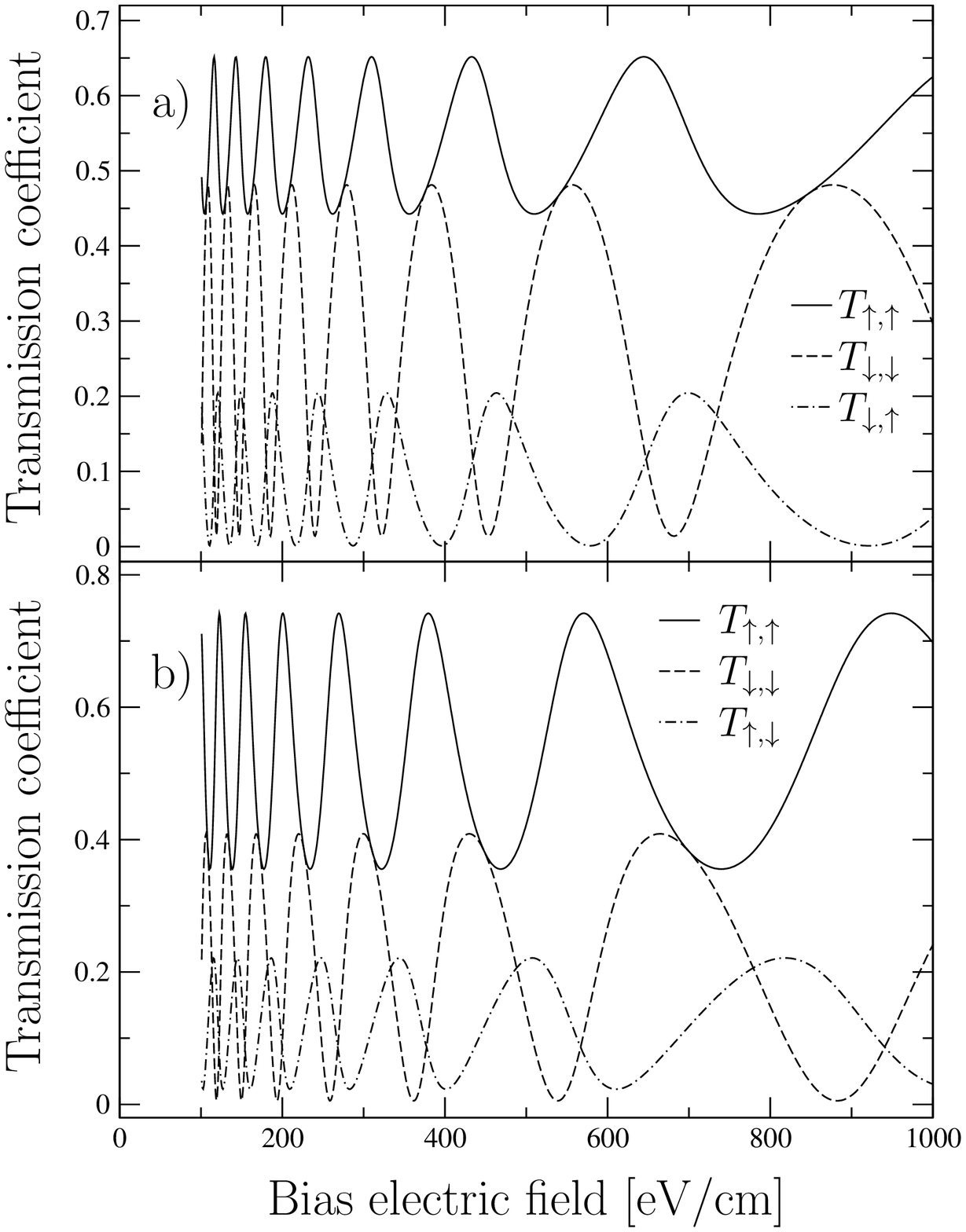}}
\caption{Transmission coefficients as functions of the bias electric field. In panel a) $T_{\uparrow, \uparrow}$, $T_{\downarrow, \downarrow}$ and $T_{\downarrow \uparrow}$ are plotted for $\delta \mathcal{E}_g /\mathcal{E}_g^* = +0.06$, in panel b) $T_{\uparrow, \uparrow}$, $T_{\downarrow, \downarrow}$ and $T_{\downarrow \uparrow}$ are plot-ed for $\delta \mathcal{E}_g/\mathcal{E}_g^* = -0.06$. The general trends discussed in fig. \ref{transmission} are preserved and, especially, one can modulate them by varying the bias electric field.}
\label{robust}
\end{figure}

An important question is how robust the results are if we change the value of the gate electric field $\mathcal{E}_g^*$ by $\mathcal{E}_g^* \pm \delta \mathcal{E}_g$, where $\frac{\delta \mathcal{E}_g}{\mathcal{E}_g^*} = \pm 0.06$. In fig. \ref{robust} we plot the transmission coefficients as function of the bias electric field: in panel a) $T_{\uparrow, \uparrow}$, $T_{\downarrow, \downarrow}$ and $T_{\downarrow \uparrow}$ are plotted for $\frac{\delta \mathcal{E}_g}{\mathcal{E}_g^*} = +0.06$, in panel b) $T_{\uparrow, \uparrow}$, $T_{\downarrow, \downarrow}$ and $T_{\downarrow \uparrow}$ are plotted for $\frac{\delta \mathcal{E}_g}{\mathcal{E}_g^*} = -0.06$. As shown there, the transmission curves are not always perfect: the minimum values of $T_{\downarrow,\downarrow}$ are not zero and their position change. However, the general trends discussed in fig. \ref{transmission} are preserved and, especially, one can modulate them by varying the bias electric field.

All of the results presented so far are valid when only a single Fourier $s$ mode propagates in the base. If more modes are allowed to mix, when $\alpha \neq \beta$ and with a magnetic field, the coefficient transmissions pattern become more complex but it is still possible to have an analytical solution by using the Sylvester Theorem for matrix-valued functions for solving the Pauli eq. Details will be given elsewhere. In ref. \cite{PPereyra1} was discussed the multichannel transmission coefficients for a FET system using this approach where the SOI is not relevant.

In summary, we combined the spin precession in the base of a spin-FET,  due to the induced spin-orbit coupling, with the analytical solution of the Pauli eq., and applied it to calculate the transfer matrix at the base. We showed that we can select the spin of the outgoing electrons to be the same as or opposite to that of the injected spin polarized electrons. More important, we can have a nearly binary square-wave transmission spin-valve effect͒ for the spin-up orientation.

\acknowledgments
The authors would like to thank Professor J. Gravinsky, for clarifying discussions. 

\bibliographystyle{eplbib}
\bibliography{Spinmosfet}

\end{document}